\documentclass[global,twocolumn]{svjour}
\usepackage{graphicx,color}
\usepackage{cite}
\usepackage{SIunits}
\usepackage{textcomp}
\journalname{Applied Physics B}
\begin{document}
	\title{A self-injection locked DBR laser for laser cooling of beryllium ions}

	\author{Steven A. King\inst{1} \and Tobias Leopold\inst{1} \and Premjith Thekkeppatt\inst{2} \and Piet O. Schmidt\inst{1,3} 
	}                     
	%
	%
	\institute{Physikalisch-Technische Bundesanstalt, 38116 Braunschweig, Germany \and International School of Photonics, Cochin University of Science and Technology, Kerala, India \and Institut f\"ur Quantenoptik, Leibniz Universit\"at Hannover, 30167 Hannover, Germany}
	\date{Received: date / Revised version: date}
	%
	\maketitle
	\begin{abstract}
		We present a simple, robust, narrow-linewidth, frequency-doubled semiconductor laser source suitable for laser cooling and repumping of $^9$Be$^+$ ions. A distributed Bragg reflector (DBR) laser diode operating at 626~nm is self-injection-locked to a frequency doubling cavity via phase-stabilised optical feedback when the laser is resonant with the cavity mode. The short-term laser instability is reduced from the MHz-level to approximately 20~kHz by the injection process, thus eliminating the need for a high-bandwidth feedback loop to suppress the otherwise troublesome high-frequency laser noise. Long-term stability of the laser frequency is achieved by feeding back to the length of the enhancement cavity utilising an electro-optic frequency comb generator to produce a beatnote with a laser that is detuned by 98~GHz. Long-term injection locking and frequency stabilisation via a wavemeter are ensured using automatic relocking algorithms. 
	\end{abstract}
	\section{Introduction}
	\label{sec:Intro}
	Owing to its simple level structure, strong laser cooling transition and non-zero nuclear spin leading to hyperfine structure, singly-ionised beryllium has found use throughout the atomic physics community for applications ranging from quantum information processing \cite{blatt_entangled_2008, gaebler_high-fidelity_2016}, atomic frequency standards \cite{bollinger_303-mhz_1991}, and as a refrigerant ion for sympathetic cooling \cite{larson_sympathetic_1986,kielpinski_sympathetic_2000,barrett_sympathetic_2003} of atomic \cite{schmidt_spectroscopy_2005, schmoger_coulomb_2015, roth_sympathetic_2005} and molecular \cite{blythe_production_2005,roth_production_2005,roth_production_2006} ion species and even (anti-)protons \cite{smorra_base_2015, bohman_sympathetic_2017} without readily accessible transitions for laser cooling.\\
	
	\begin{figure}
		\resizebox{\columnwidth}{!}{%
			\includegraphics{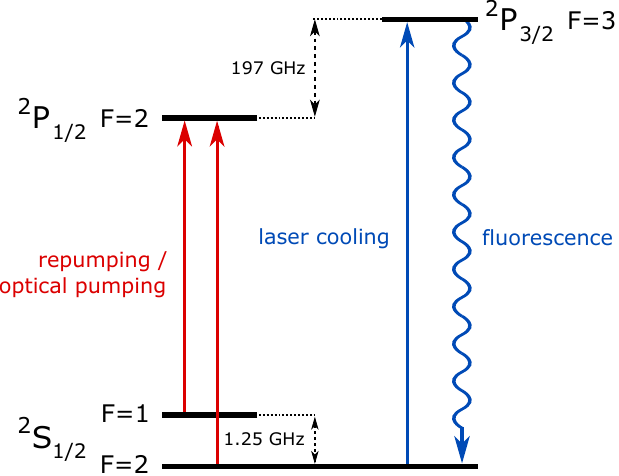}
		}
		\caption{Partial term scheme of $^9$Be$^+$ (not to scale), showing the transitions near 313~nm used for laser cooling (blue) and repumping/optical pumping (red). Spontaneously emitted photons from the $^2$P levels can be detected using a photomultiplier or CCD camera. Other hyperfine levels in the $^2$P manifold have been omitted for clarity. The fine structure splitting of almost 200~GHz between the $^2$P levels necessitates separate laser sources for cooling and repumping.}
		\label{fig:Be}       
	\end{figure}
	
	A simplified term scheme of $^9$Be$^+$ is shown in figure~\ref{fig:Be}. Many approaches have been taken in order to produce light at 313~nm, corresponding to the wavelengths of the strong $^2$S$_{1/2}$ $\rightarrow$ $^2$P$_{1/2}$ and $^2$S$_{1/2}$ $\rightarrow$ $^2$P$_{3/2}$ transitions used for laser cooling and repumping, including frequency doubling of dye lasers \cite{bollinger_hyperfine_1985}, sum-frequency generation of Nd:YAG and Ti:sapphire lasers followed by frequency doubling \cite{schnitzler_all-solid-state_2002}, sum frequency generation of two near infra-red lasers followed by frequency doubling \cite{wilson_750-mw_2011}, and second- \cite{cozijn_laser_2013, ball_high-power_2013}, third- \cite{carollo_third-harmonic-generation_2017} or fifth- \cite{vasilyev_compact_2011} harmonic generation of diode lasers. Previous semiconductor laser systems that have produced 626~nm directly required cooling of the laser diodes to around -30$^{\circ}$C \cite{cozijn_laser_2013,ball_high-power_2013}, which can be a technical challenge.\\
	 In this paper, we demonstrate a laser system based on a low-cost, frequency-doubled distributed Bragg reflector (DBR) ridge waveguide laser diode to produce radiation at 313~nm. Rather than using a high-bandwidth pre-stabilisation of the 626~nm laser to an external reference to narrow its linewidth prior to frequency doubling, we exploit optical feedback from the frequency doubling enhancement cavity itself to narrow the laser linewidth. This greatly simplifies the necessary locking setup, and both maximises the achievable UV power after doubling and reduces its amplitude noise.

	\section{Laser system}
	\label{sec:Laser}
	
	\begin{figure}
	\resizebox{\columnwidth}{!}{%
	\includegraphics{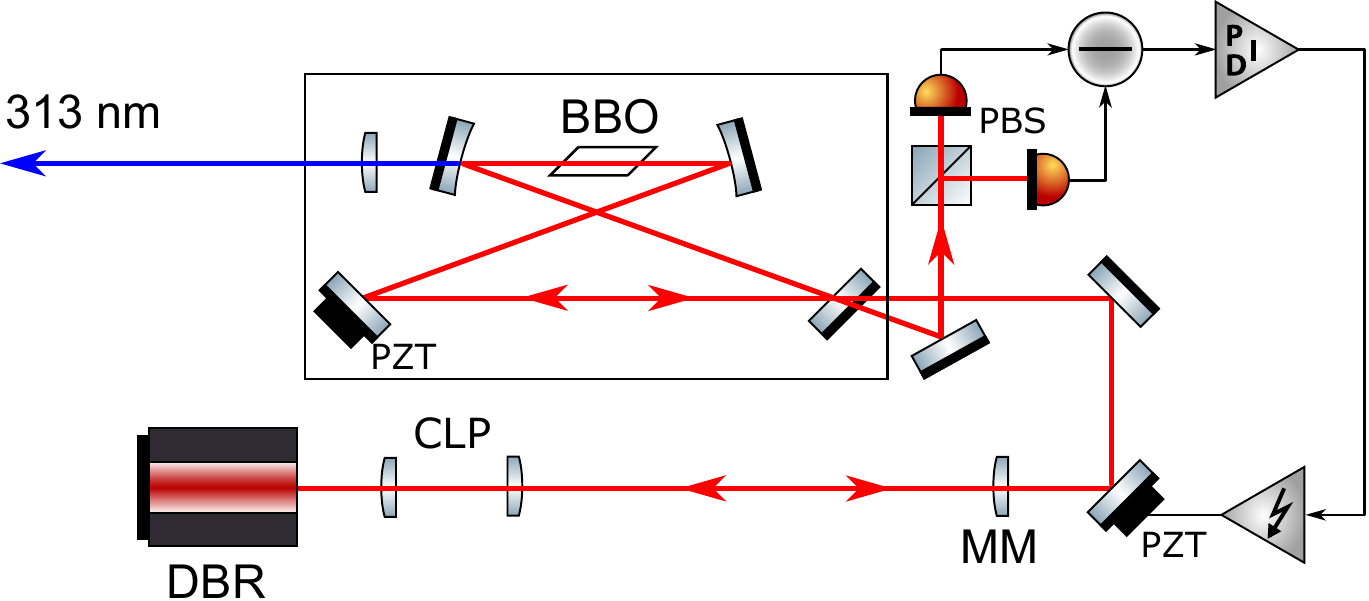}
		}
	\caption{Simplified schematic showing an overview of the laser system (colour online), with the laser self-injected using the backward-circulating light from the doubling cavity. The bidirectional parts of the beam path are indicated by opposing arrows. CLP: cylindrical lens pair, PZT: piezoelectric transducer, MM: mode matching lens, PBS: polarising beamsplitter.}
	\label{fig:Laser}       
	\end{figure}
	
\begin{figure}
		\resizebox{\columnwidth}{!}{%
			\includegraphics{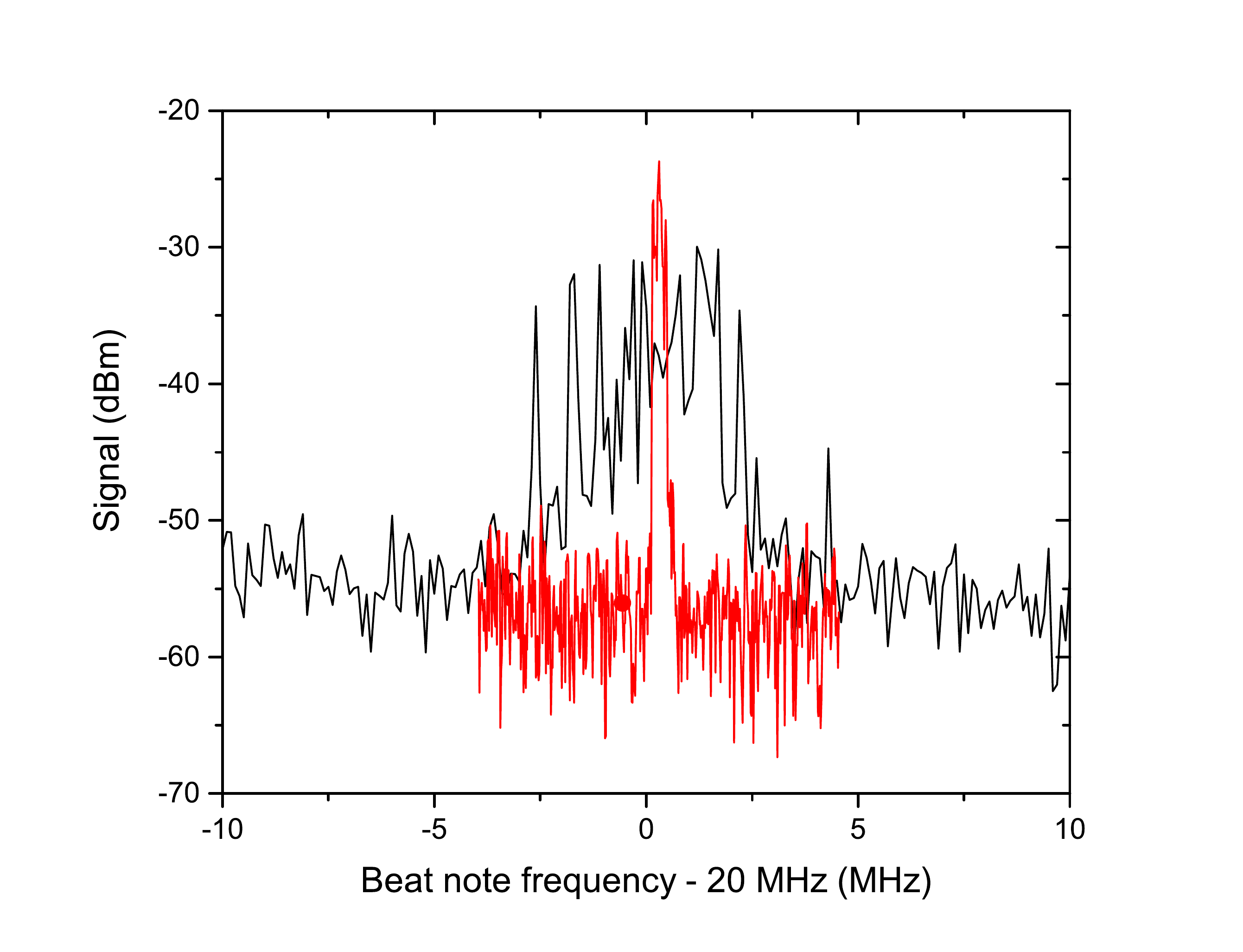}
		}
		\caption{Beat note between a narrow linewidth iodine-stabilised 626~nm laser and the DBR laser used in this work for the cases of the free-running DBR laser (black) and after weak injection locking (red). There is a marked reduction in the high frequency noise after the injection lock.}
		\label{fig:Beat}       
	\end{figure}

	A simplified schematic of the laser system is shown in figure~\ref{fig:Laser}. The laser used in this work is a distributed Bragg reflector (DBR) ridge waveguide laser diode produced by the Ferdinand-Braun Institute \cite{blume_monolithic_2013}. At a temperature of 5.6$^\circ$C and drive current of 130~mA, approximately 26~mW of light is produced at 626.395~nm. At the maximum injection current of 150~mA, 42~mW of light is emitted. An integrated Peltier thermoelectric cooler and hermetic sealing of the diode case allows cooling below the dew point of the laboratory in a single stage without any additional precautions. The steady-state current drawn by the Peltier cooler is approximately 400~mA. A passive shield covering the laser decouples it from fluctuations of the room temperature. The free-running laser frequency is stable to around 50~MHz over the course of a day, and therefore the temperature fluctuations of the diode chip are estimated to be at the millikelvin level on these timescales based on a measured temperature sensitivity of 30~MHz/mK. The laser is driven by a home-built low-noise current supply based on the Libbrecht-Hall design \cite{libbrecht_lownoise_1993}, but even so a free-running linewidth of approximately 2~MHz is observed.\\	
	The divergent laser output is collimated using an aspheric lens with 4.51~mm focal length and large numerical aperture of 0.54 to prevent clipping of the beam. The 6:1 aspect ratio of the output beam is corrected using a pair of cylindrical lenses. In the initial setup, the laser then passed through a Faraday isolator. The red light is delivered to a mechanically monolithic enhancement cavity \cite{hannig_highly_2018}, where it is converted to 313~nm using a 10~mm-long beta-barium borate (BBO) crystal cut at the phase matching angle of 38.4\degree. To reduce loss due to reflection, the input and output facets of the crystal are cut at 59\degree, corresponding to Brewster's angle for the 626~nm light. The cavity is sealed and continuously purged by a low flow of oxygen gas, which has been shown to prevent degradation of the intracavity elements in the presence of UV light \cite{gangloff_preventing_2015, kunz_experimentation_2000, cooper_cavity_2018, brandstatter_integrated_2013,pereira_dynamic_2009}.\\
	Concave mirrors with a radius of curvature of 50~mm are used to produce a waist of \unit{29}{\micro\meter} inside the crystal. 0.4~mW of UV radiation is generated from 23~mW of input 626~nm light. Approximately 70\% of the input light is coupled into the cavity. 
	If the doubling cavity length is stabilised to the input laser via the H\"ansch-Couillaud technique, the locking bandwidth is ultimately limited by the lowest frequency mechanical resonance of the piezoelectric transducer (PZT) attached to an intracavity mirror used for the feedback, which lies at around 50~kHz. This level of bandwidth is not sufficiently high to track the high frequency laser noise, and fast fluctuations of around 50\% in the output UV power are observed with the cavity length stabilised since the laser and doubling cavity have similar linewidths.\\	
	To suppress the high-frequency noise on the laser diode and hence improve the coupling of the laser to the cavity, a high-bandwidth feedback loop to a stable external reference would normally be required. We have taken another approach based on self-injection locking of the laser \cite{dahmani_frequency_1987}, a technique which has been successfully employed to passively narrow the short-term linewidth of diode lasers all the way to the hertz level \cite{zhao_high-finesse_2012}. By using feedback from the enhancement cavity itelf \cite{hemmerich_second-harmonic_1990}, we do not require an additional delay line \cite{lin_long-external-cavity_2012} or folding back of the transmission of an additional optical cavity \cite{zhao_high-finesse_2012, lewoczko-adamczyk_ultra-narrow_2015}. If the optical isolation between the doubling cavity and the laser is removed, feedback into the laser diode is observed when the laser is resonant with one of the cavity modes. This feedback is somewhat surprising as the cavity is in a bow-tie (travelling wave) configuration and so the leaked/reflected light should not be counterpropagating to the input beam. However, it was observed that there is a backward-travelling wave in the cavity with a power level approximately $2\times 10^{-5}$ of the forward-travelling wave. This wave is likely seeded by scatter from the surface of one of the intracavity elements, most likely the facet of the doubling crystal \cite{hemmerich_second-harmonic_1990}. It is unlikely to be seeded by backscattering from an induced grating in the BBO crystal caused by the high laser intensities there, as the ratio between the forward- and backward-circulating powers was found to be independent of input power up to the maximum power output of the laser.\\ 	
	This counterpropagating beam provides frequency-selective feedback of approximately -47~dB to the laser diode, which is at the lower end of the range required for stable single-mode operation \cite{lin_long-external-cavity_2012}. After injection locking, the linewidth of the laser diode was observed to significantly narrow as shown in figure~\ref{fig:Beat}. The capture range for the injection with this level of feedback is approximately 130~MHz.\\	
	Various environmental effects such as air currents and vibrations cause variation in the phase of the feedback into the laser diode, leading to slow fluctuations of the locked laser frequency from the peak of the cavity transmission fringe. To correct for this, the H\"ansch-Couillaud error signal can be used. This signal is sent to a digital PID controller, and electronic feedback is applied to a PZT connected to a mirror in the optical feedback path. As such, the laser is locked to the peak of the doubling cavity transmission fringe and the maximum UV output power is maintained.\\	
	It was found that weak injection still occurs even when a Faraday isolator with 47~dB of isolation is placed in front of the laser, observed as the laser weakly tracking the doubling cavity resonance even without electrical feedback, but without any suppression of the high frequency laser noise. This means that a further isolation stage would be required if the laser was to be completely protected from injection by the backwards-travelling beam. A major advantage of exploiting the injection in this configuration is therefore the avoidance of the significant transmission loss through a dual-stage isolator, thus allowing full use of the relatively low laser power and maximising the achievable UV power.\\	
	The UV beam is split into two paths, each containing a double-pass acousto-optic modulator (AOM) to produce a frequency difference of 1.25~GHz between the two beams. This allows the same laser to be used for both repumping and optical pumping purposes. The two beams are then recombined and coupled into an FC/APC connectorised, hydrogen-loaded, UV-cured large mode area (LMA) optical fibre \cite{colombe_single-mode_2014, marciniak_towards_2017} to transfer the light to the ion trap setup. A transmission efficiency of over 60\% is achieved for a fibre of 1.5~m length when tight bend radii are avoided.
	
	\section{Increasing the feedback power}
	\label{sec:BrewsterFeedback}
	
\begin{figure}
		\resizebox{\columnwidth}{!}{%
			\includegraphics{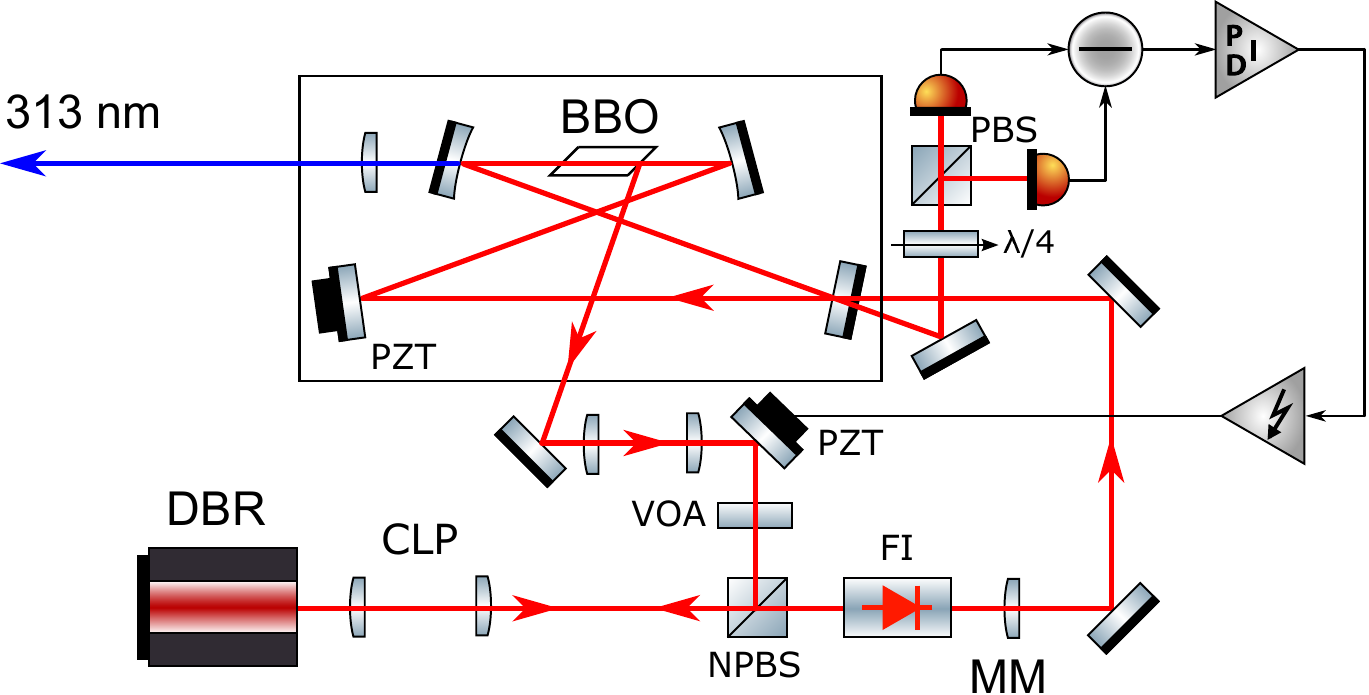}
		}
		\caption{Simplified schematic showing an overview of the laser system (colour online), after the laser was self-injected using the reflection from the Brewster-cut facet of the BBO crystal. The bidirectional parts of the beam path are indicated by opposing arrows. FI: Faraday isolator, VOA: variable optical attenuator, NPBS: non-polarising beamsplitter.}
		\label{fig:Laser2}       
	\end{figure}

	To increase the robustness and capture range of the injection lock, it was necessary to increase the feedback fraction into the laser diode. To this end, the reflection from the Brewster-angled facet of the BBO crystal was extracted via a window in the enhancement cavity housing \cite{hannig_highly_2018} and was folded back into the laser diode via a non-polarising beamsplitter with a reflectivity of 2\% as shown in figure~\ref{fig:Laser2}.\\
	When the laser is resonant with the TEM$_{\textrm{00}}$ mode of the cavity and the circulating power is maximised, the minimal level of reflection from the Brewster-cut surface with optimised laser polarisation is approximately 0.75~mW when 23~mW is incident on the cavity. Care was taken to ensure no other optics were normal to the laser beam, which could lead to competition between various sources of optical feedback. In addition, the single-stage optical isolator between the doubling cavity and the laser was restored, otherwise competition was observed from this feedback path. Two lenses were used in the feedback path from the Brewster reflection to improve the spatial mode matching between the forward and backward travelling beams, but the lenses were deliberately left unoptimised in order to prevent further seeding of the backward-travelling wave in the doubling cavity which could then become a competing feedback path despite the presence of the optical isolator. The astigmatism in the reflection from the Brewster surface was also left uncorrected for this reason. Alignment of the feedback beam into the diode was optimised by observing the transmission fringes of the cavity as its length was swept by several free spectral ranges (FSR). As the feedback alignment improves, the fringes are observed to broaden due to the increased capture range. A variable optical attenuator in the feedback path allowed variation of the feedback level, and also provided some isolation to further prevent seeding of the backward-travelling wave in the doubling cavity. With the alignment optimised and an attenuation of 10~dB, the capture range of the injection lock was increased to 210~MHz. Because of the imperfect matching back into the diode, it is difficult to quantify the feedback fraction relative to the emitted light. The power incident on the laser collimating lens was measured to be \unit{1.2}{\micro\watt}, corresponding to a power ratio of -43~dB to the emitted power of 26~mW. It was chosen to operate with this level of feedback, despite the fact that a greatly increased capture range of 800~MHz is achieved if the attenuator is completely removed. In this case, the returning power ratio is approximately $-33$~dB. At such a high feedback level, the laser is also able to stably operate at two additional frequencies at $\pm22$~GHz from the free-running state which could correspond to side modes of the diode.
	
	\section{Stabilising the laser wavelength}
	\label{sec:EOMComb}
	
\begin{figure}
		\resizebox{\columnwidth}{!}{%
			\includegraphics{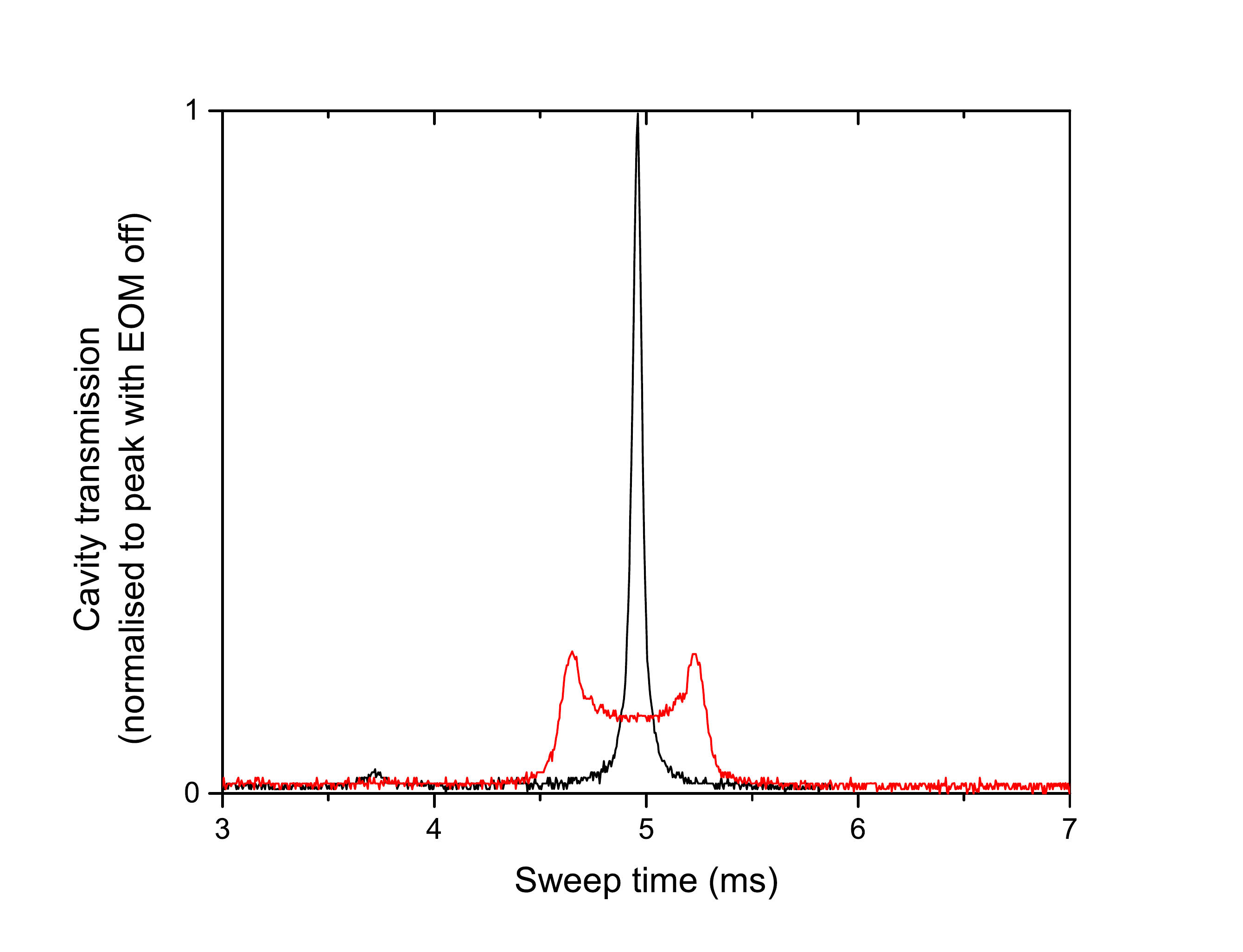}
		}
		\caption{Transmission profile of the EOM cavity obtained by scanning the cavity length with the EOM turned off and on. The modulation depth was determined by comparison of the width of the FM lineshape to the FSR of the cavity.}
		\label{fig:Transmission}       
	\end{figure}

\begin{figure}
		\resizebox{\columnwidth}{!}{%
			\includegraphics{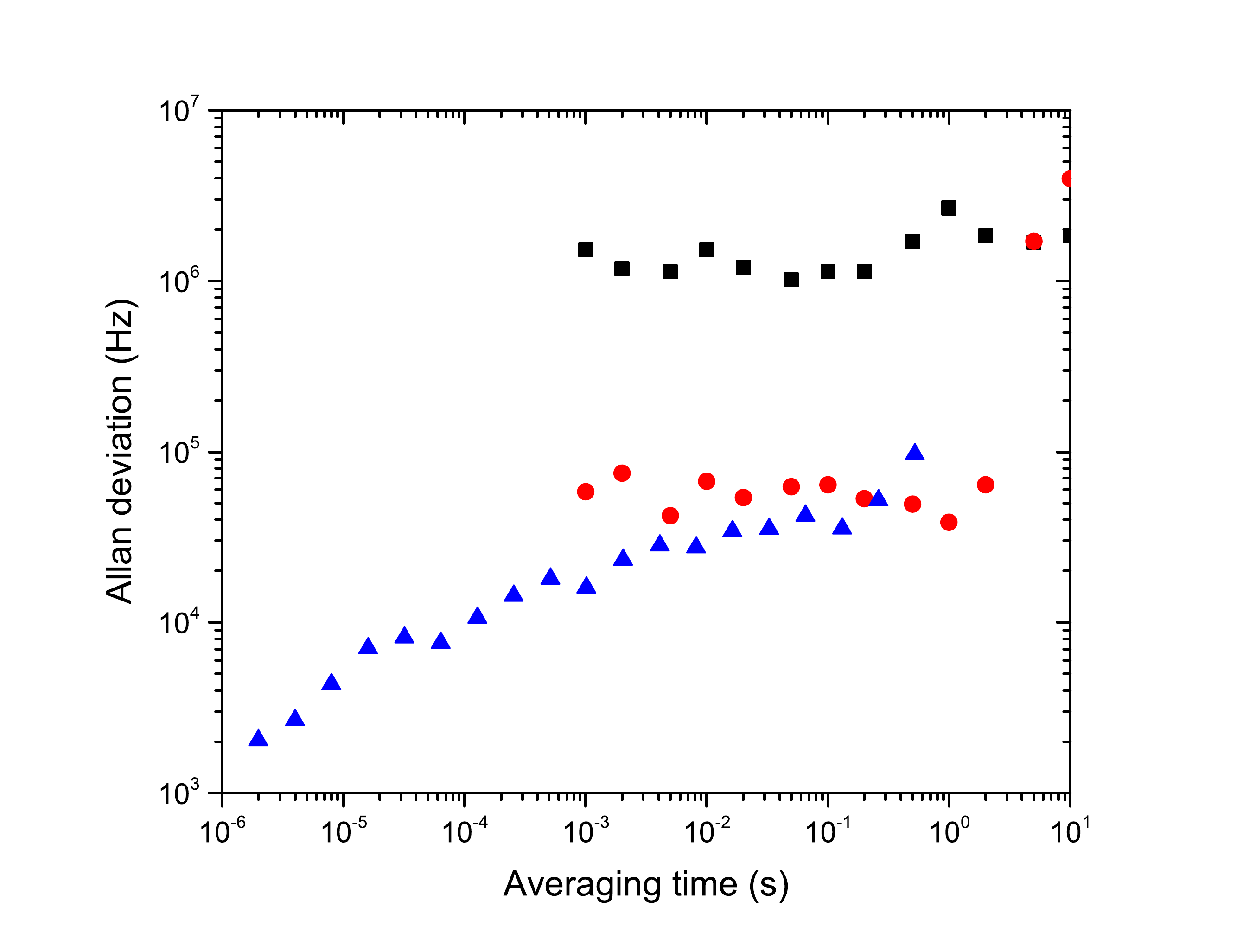}
		}
		\caption{Allan deviation of the DBR laser frequency in the case of the laser free-running with no injection (black squares), injection from the backward-circulating light (red circles), and attenuated injection from the reflection from the BBO crystal facet (blue triangles). The stabilities in the cases of the free-running laser and with injection from the backward-circulating beam were determined using the beat note with the SFG laser, and the stability with feedback from the Brewster facet was determined using the side-of-fringe technique with a stable analysis cavity.}
		\label{fig:Adev}       
	\end{figure}

\begin{figure}
		\resizebox{\columnwidth}{!}{%
			\includegraphics{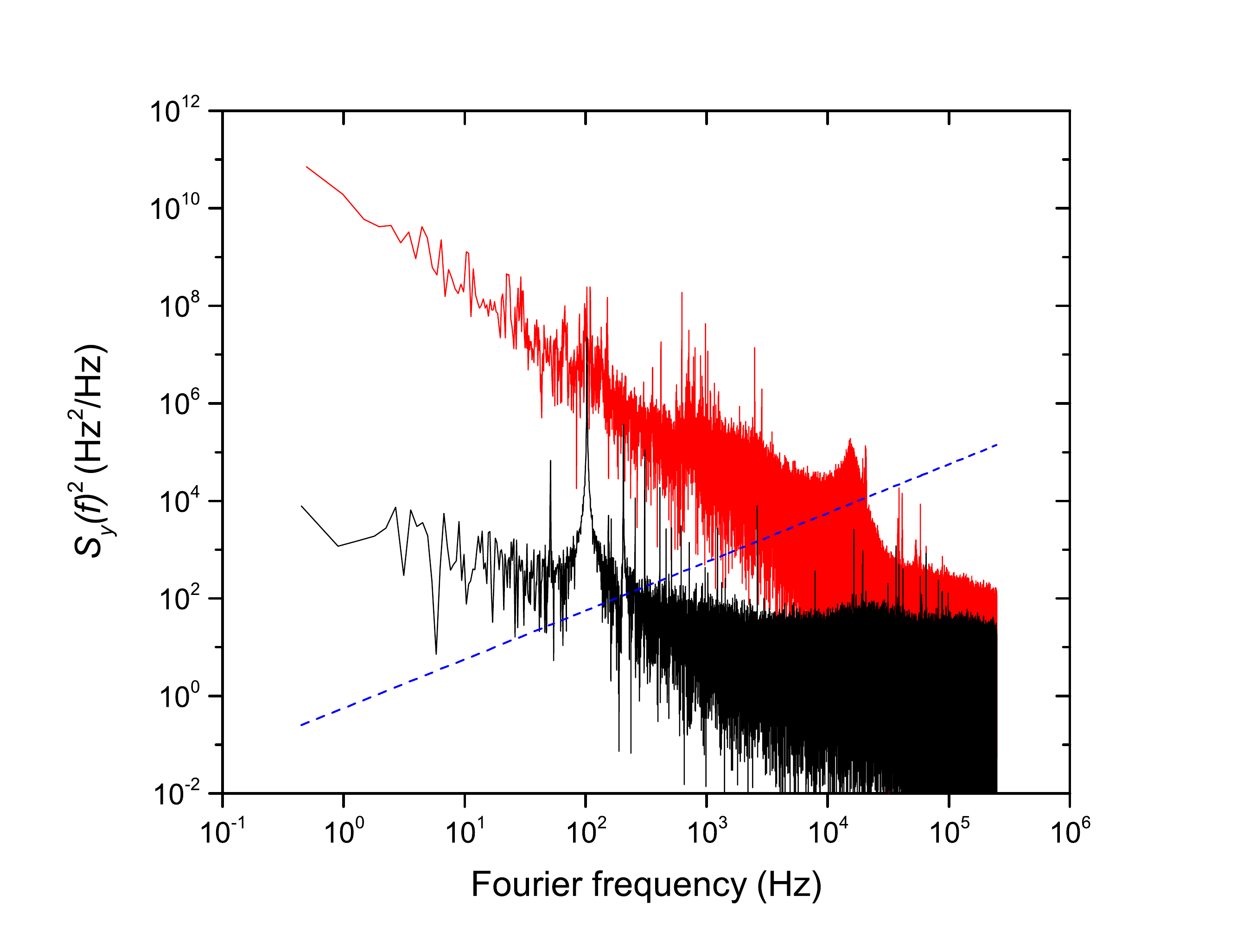}
		}
		\caption{Red: Power spectral density (PSD) $S_y(f)^2$ of the laser frequency noise after injection from the attenuated reflection from the Brewster-cut BBO crystal facet, as observed using the side-of-fringe technique. Black: measurement noise floor. The features at 50~Hz and its harmonics arise from the mains frequency, with the pronounced feature at 100~Hz likely caused by unsuppressed scatter from the room lights. Blue: $\beta$-separation line, used for estimating the linewidth of the locked laser.}
		\label{fig:PSD}       
	\end{figure}

	The doubling cavity is mechanically monolithic, which leads to greatly reduced vibration sensitivity and good short-term stability of the injection-locked laser frequency. However, drifts in the length of the cavity can still occur due to variations in laboratory temperature, drift of the PZT element, intracavity laser power, and changes in pressure due to the oxygen purging of the cavity. We observe a drift in the injection-locked laser frequency of a few hundred kilohertz per second (corresponding to a doubling cavity length drift of a few hundred pm/s), which would greatly reduce its operational usefulness. To eliminate this drift, it is necessary to provide slow feedback to the length of the doubling cavity via feedback to the PZT on the intracavity mirror. As the DBR laser does not have enough spare output power for (for example) stabilising to an atomic reference via saturated absorption spectroscopy, and building a highly stable optical reference cavity is prohibatively expensive, it was decided to lock the laser to another 626~nm laser available in the laboratory, used for laser cooling of Be$^+$ ions on the $^2$S$_{1/2}$ $\rightarrow$ $^2$P$_{3/2}$ transition. This light is produced by sum frequency generation (SFG) of two infra-red fibre lasers and is locked to an absorption line in molecular iodine, providing excellent long-term stability. The two lasers are separated in frequency by approximately 98~GHz as the DBR laser is intended to be used for repumping and optical pumping of the ion via the $^2$P$_{1/2}$ state.\\ 
	\begin{figure}
		\resizebox{\columnwidth}{!}{%
			\includegraphics{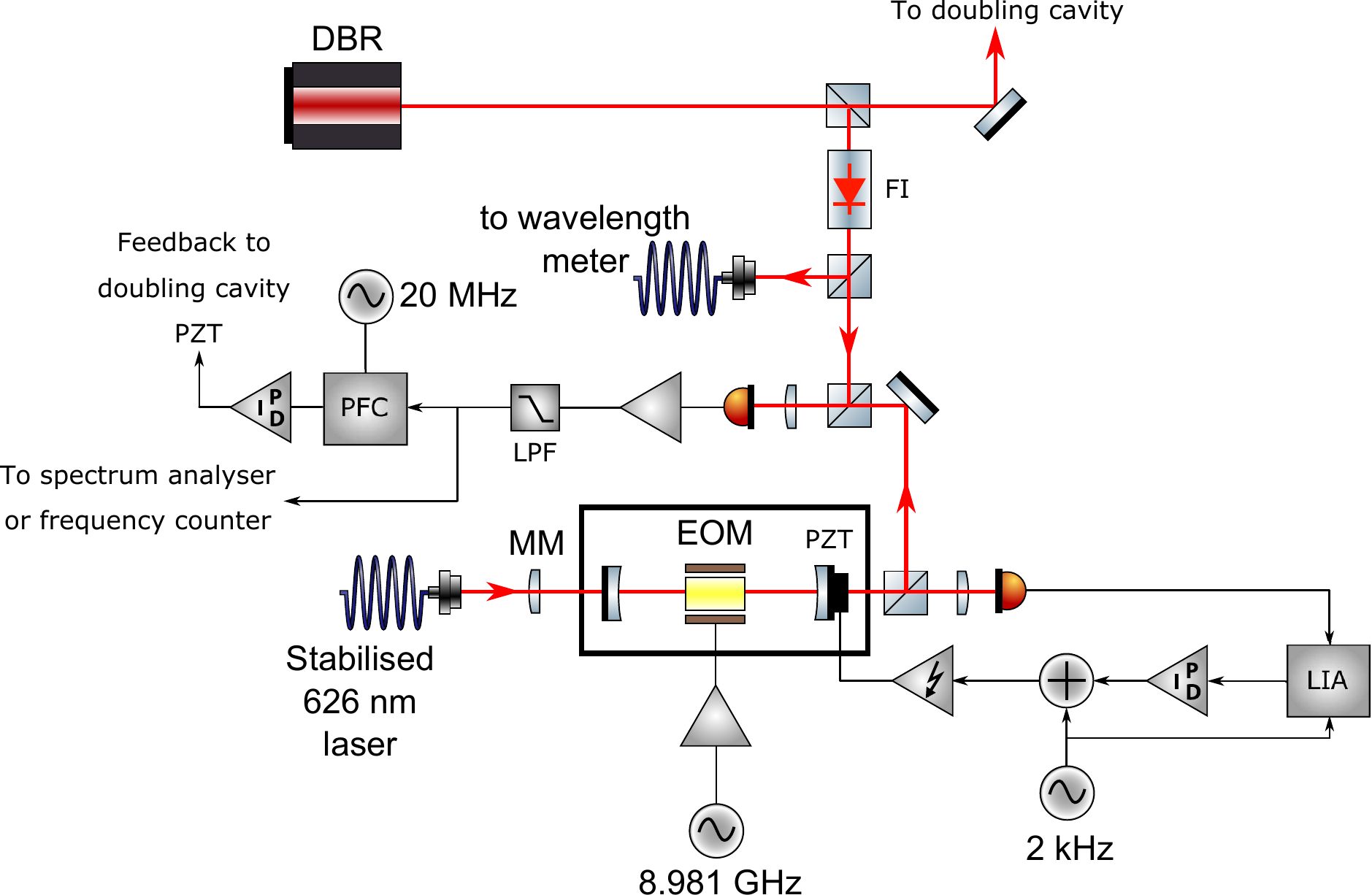}
		}
		\caption{Simplified schematic showing an overview of the layout for producing a beat note between the two 626~nm lasers using an electro-optic frequency comb generator, and optionally feeding back to the laser frequency via the doubling cavity PZT. LIA: lock-in amplifer, PFC: phase-frequency comparator, MM: mode matching lens, LPF: low-pass filter.}
		\label{fig:EOMCavity}       
	\end{figure}
	  Producing a beat note between the two lasers is non-trivial as the offset frequency is far too high for conventional electronics to detect and process. To bridge this large frequency gap, a sample of the SFG laser is split off and its spectrum broadened. An electro-optic phase modulator (EOM) operating at 8.981~GHz is placed within an optical cavity, as shown in figure~\ref{fig:EOMCavity}, formed from two concave mirrors with 600~mm radius of curvature and transmission of 1\% at 626~nm. The finesse of the cavity is approximately 83 with the EOM present, which is likely limited by transmission loss and the significant beam profile distortion introduced by the EOM. This results in the cavity being significantly undercoupled, with a coupling fraction of only 7.5\% and peak transmission of only 1.6\% of the input light. When the FSR of the cavity is equal to a sub-harmonic of the EOM drive frequency, the phase modulation sidebands produced by the EOM are also resonant with the cavity. In this case, a FSR of approximately 1~GHz was chosen, corresponding to a mirror separation of 114~mm with the EOM present. Repeated passes through the EOM transfers the laser power into higher and higher order sidebands until the light leaks out of the cavity. In this manner, frequency combs spanning several THz have been produced \cite{kourogi_wide-span_1993}, for applications including frequency chains \cite{margolis_3_2001}. Further broadening is generally limited by dispersive effects in the EOM crystal.\\	
	To stabilise the EOM cavity length to the SFG laser and maintain the maximum output power, a dither lock was employed. Using a PZT attached to one of the cavity mirrors, the cavity length was dithered at a frequency of 2~kHz. The laser power transmission through the cavity is monitored by a photodiode and sent to a lock-in amplifier, where it was compared to the dither signal. Feedback was applied to the cavity length by another digital PID controller via the same PZT connected to one of the cavity mirrors. Because of the FM lineshape of the cavity with the EOM turned on, the transmission of the cavity on resonance was reduced by a further order of magnitude as shown in figure~\ref{fig:Transmission}.\\
	The transmitted beam from the cavity was overlapped with \unit{300}{\micro \watt} of light from the DBR laser and sent to a high bandwidth photodetector as shown in figure~\ref{fig:EOMCavity}, resulting in a beatnote with the nearest comb mode. In this case, this was the eleventh order sideband of the EOM at 98.79~GHz from the carrier. The power transfer into the relevant comb tooth can be approximated by \cite{kourogi_wide-span_1993}:
	\begin{equation}
	\frac{P_k}{P_{\textrm{in}}} = \eta \left(\frac{\pi}{2 \beta F}\right)^2 \textrm{exp} \left( - \frac{\left| k \right| \pi }{\beta F}\right)
	\end{equation}
	where $\eta$ is the transmission fraction through the optical cavity, $\beta$ is the modulation index, $F$ is the cavity finesse and $k$ is the sideband order. For sideband order $k=11$ and a cavity finesse of 83, a maximum power transfer of 0.11\% occurs for $\beta = 0.21$ (neglecting cavity transmission loss). Under these conditions, and with 2.5~mW of 626~nm light at the input to the cavity, we would predict an output power of 4~nW in the 11th order sideband when considering the cavity transmission loss. We observe a signal-to-noise ratio (SNR) of 30~dB in the beatnote with a 30~kHz resolution bandwidth, as shown in figure~\ref{fig:Beat}. The SNR could likely be improved by employing an additional filter cavity at the the output, allowing the selection of only the comb mode of interest and hence reducing the shot noise caused by the unused comb modes.\\		
	Using a Stanford SR620 frequency counter with a gate time of 1~ms, the stability of the beat note was measured before and after the injection locking using the backwards circulating light. It can be seen in figure~\ref{fig:Adev} that the DBR laser stability was improved by more than an order of magnitude on timescales between 1~ms and 1~s by the weak optical feedback.\\ 
	To measure the laser stability on shorter timescales, and to remove contributions due to the stability of the SFG laser, a sample of the laser light was delivered to a stable optical cavity. The cavity converts frequency fluctuations of the laser to amplitude noise observable on a fast photodiode when the laser is tuned to the half-maximum of the transmission fringe. The linewidth of the cavity was measured to be 2.6(2)~MHz at 626~nm by tuning the laser across the resonance using an acousto-optic modulator. Data was collected for a period of 2~s at a sample rate of 500~kSa/s using a data acquisition system based on an FPGA (STEMlab, formerly Red Pitaya). Before injection locking, the laser noise caused fluctuations much greater than the cavity fringe width, preventing measurements of the free-running laser noise. After injection, it was easily possible to tune the laser to the half-maximum point of the analyser cavity resonance by applying a voltage to the PZT inside the doubling cavity. The Allan deviation of the signal transmitted through the cavity is shown in figure~\ref{fig:Adev}. The laser stability was improved by two orders of magnitude compared to the free-running case at an averaging time of 1~ms, and is better than 20~kHz out to a timescale of 1~ms. The improved performance compared to injection via the backwards circulating light is a consequence of the stronger optical feedback.
	The power spectral density of the transmitted signal is shown in figure~\ref{fig:PSD}. There is a clear $1/f$ (flicker frequency) behaviour, which has been observed with such systems \cite{breant_ultra-narrow_1989}. As discussed in \cite{lin_long-external-cavity_2012}, at sufficiently high Fourier frequencies the noise PSD becomes independent of frequency, corresponding to the intrinsic Lorentzian linewidth of the laser. From this data, no obvious plateau is reached. The noise power spectral density of 16~Hz$^2/$Hz at the highest measured Fourier frequency allows us to place an upper limit of 50~Hz on the intrinsic laser linewidth. The Schawlow-Townes limit for an effective extended cavity diode laser with resonator length of 1~m and feedback fraction of -43~dB would be approximately 100~Hz, which lends credence to the measured value. 
	It is difficult to assign a linewidth to the laser because of the $1/f$ characteristic of the noise \cite{matei_1.5_2017}, but by finding the intercept between the noise power spectral density and the $\beta$-separation line, an approximate linewidth can be estimated \cite{di_domenico_simple_2010}. This intercept occurs at a frequency of 9.3~kHz, indicating a linewidth of approximately 51~kHz at our measurement time of 2~s for the case of a strongly injected diode.
	As shown in figure~\ref{fig:EOMCavity}, the lasers separated by almost 100~GHz were then locked together by demodulating the beat note between the EOM frequency comb and the DBR laser and using a phase-frequency comparator to produce an error signal. Low-bandwidth feedback was then applied to the PZT inside the doubling cavity such that slow drifts in its length were compensated. The locked laser can then be easily tuned in frequency by changing the demodulation frequency, until the capture range of the injection lock is exceeded.

	\section{Lock to wavemeter and automatic recovery of injection lock}
	\label{sec:Wavemeter}
	
	A simpler (but more expensive) method of stabilising the laser wavelength is to use a high-accuracy wavemeter \cite{saleh_frequency_2015,couturier_laser_2018,kobtsev_long-term_2007}. The wavemeter reading is processed by software, and the wavelength compared to a pre-determined setpoint. Feedback can then be applied to the length of the doubling cavity via the PZT by amplifying the output of a digital-to-analog converter (DAC) attached to the control computer. This method has the advantage that the laser frequency can easily be tuned and monitored as part of routine experimental diagnostics, but will result in lower mid-term performance due to the (at best) MHz-level precision of the wavemeter reading and possible long-term drifts in accuracy if the wavemeter is not frequently calibrated. This level of performance is generally sufficient for repumping purposes, however.\\	
	To improve the long-term robustness of the system, it is desirable to have the status of the injection lock permanently monitored and automatically recovered if necessary. This could be required in the case of a large shock to the optical table, or the feedback PZT reaching the end of its travel. To achieve this, the waste red light transmitted by the cavity output coupler is monitored by a photodiode, whose output is amplified to produce a TTL-compatible signal which is read by a digital input channel on the same card that tunes the doubling cavity length. In the event that the injection lock is lost, the monitoring software applies a linear ramp to the doubling cavity PZT until transmission is once again observed. At this point, the feedback phase stabilisation servo automatically re-captures and the wavelength stabilisation servo is re-engaged.

	\section{Conclusions}
	\label{sec:Conclusions}
	We have presented a narrow-linewidth, robust laser system for laser cooling and repumping of Be$^+$ ions. The simplicity of the system makes it attractive for long-term robust operation, and the self-injection locking of the laser diode allows the realisation of a narrow linewidth and low amplitude modulation on the generated UV light without the need for a feedback loop with MHz-level bandwidth. Owing to the excellent temperature stability of the laser diode, wide capture range and automatic restoration of the injection lock, the laser is able to operate for extended periods without user intervention. Long-term however, it may be desirable to detect and avoid impending mode-hops via observation of increasing amplitude noise on the laser light \cite{heumier_detecting_1993,chiow_extended-cavity_2007}, or occasional monitoring of the free-running laser wavelength and implementation of a slow feed-forward onto the laser current.  In the case where feedback is induced by backward-circulating light from the doubling cavity, the lack of optical isolation also allows access to the full laser power, which could be useful for laser systems with limited output power. It is also worth noting that it would be possible to injection lock a high power slave laser or seed a tapered amplifier with the stabilised light, for situations where a higher UV output power is required. It would be possible to produce a detuning of up to 20~GHz to the master laser using acousto- or electro-optics, allowing the driving of stimulated Raman transitions with low levels of off-resonant scattering \cite{hemmerling_single_2011}. In addition, this work could find applications for pre-stabilisation of noisy low-cost red or IR lasers for laser cooling and clock interrogation in transportable optical clock systems, where frequency doubling using enhancement cavities is required to reach sufficiently high powers at the desired wavelength \cite{ludlow_optical_2015, cao_compact_2017, koller_transportable_2017}.
	
	\section{Acknowledgements}
	The authors gratefully thank Uwe Sterr and Dirk Piester for generously loaning equipment to make this project possible, Helen Margolis for her advice regarding the EOM frequency comb, and thank Lennart Pelzer and Mariia Stepanova for their contributions to the experiment. SAK acknowledges support by the Alexander von Humboldt Foundation, and PT was supported by the IP@Leibniz programme of the Leibniz Universit\"at Hannover. We acknowledge support from the Deutsche Forschungsgemeinschaft through SCHM2678/5-1.

\end{document}